\documentclass[oneside,english]{elsart}
\usepackage[T1]{fontenc}
\usepackage[latin9]{inputenc}
\usepackage{float}
\usepackage{graphicx}
\usepackage{amssymb}

\makeatletter

\providecommand{\tabularnewline}{\\}

\makeatother

\usepackage{babel}

\begin{document}
\begin{frontmatter}

\title{Fortran 90 implementation of the Hartree-Fock approach within the
CNDO/2 and INDO models}

\author{Sridhar Sahu$^{1}$, Alok Shukla$^{2}$}

\address{Department of Physics, Indian Institute of Technology, Bombay, Powai,
Mumbai 400076, INDIA}

\thanks{e-mail: sridhar@phy.iitb.ac.in}

\thanks{Author to whom all the correspondence should be addressed. e-mail: shukla@phy.iitb.ac.in}

\begin{abstract}
Despite the tremendous advances made by the \emph{ab initio} theory
of electronic structure of atoms and molecules, its applications are
still not possible for very large systems. Therefore, semi-empirical
model Hamiltonians based on the zero-differential overlap (ZDO) approach
such as the Pariser-Parr-Pople, CNDO, INDO, etc. provide attractive,
and computationally tractable, alternatives to the \emph{ab initio}
treatment of large systems. In this paper we describe a Fortran 90
computer program developed by us, that uses CNDO/2 and INDO methods
to solve Hartree-Fock(HF) equation for molecular systems. The INDO
method can be used for the molecules containing the first-row atoms,
while the CNDO/2 method is applicable to those containing both the
first-, and the second-row, atoms. We have paid particular attention
to computational efficiency while developing the code, and, therefore,
it allows us to perform calculations on large molecules such as C$_{60}$
on small computers within a matter of seconds. Besides being able
to compute the molecular orbitals and total energies, our code is
also able to compute properties such as the electric dipole moment,
Mulliken population analysis, and linear optical absorption spectrum
of the system. We also demonstrate how the program can be used to
compute the total energy per unit cell of a polymer. The applications
presented in this paper include small organic and inorganic molecules,
fullerene C$_{60}$, and model polymeric systems, \emph{viz.}, chains
containing alternating boron and nitrogen atoms (BN chain), and carbon
atoms (C chain). 
\end{abstract}
\end{frontmatter}
\textbf{Program Summary} \\
 \emph{Title of program:} cindo.x \\
 \emph{Catalogue Identifier:} \\
 \emph{Program summary URL:} \\
 \emph{Program obtainable from:} CPC Program Library, Queen's University
of Belfast, N. Ireland \\
 \emph{Distribution format:} tar.gz\\
 \emph{Computers :} PC's/Linux\\
\emph{Linux Distribution:} Code was developed and tested on various
recent versions of Fedora including Fedora 9 (kernel version 2.6.25-14)\\
 \emph{Programming language used:} Fortran 90\\
\emph{Compilers used:} Program has been tested with Intel Fortran
Compiler (non-commercial version 10.1) and gfortran compiler (gcc
version 4.3.0) with optimization option -O. \emph{}\\
\emph{Libraries needed: }This program needs to link with LAPACK/BLAS
libraries compiled with the same compiler as the program. For the
Intel Fortran Compiler\emph{ we used the ACML library version 3.6.0,
while for gfortran compiler we used the libraries supplied with the
Fedora distribution.}\\
\emph{Number of bytes in distributed program, including test data,
etc.:} size of the tar file ...... bytes\\
 \emph{Number of lines in distributed program, including test data,
etc.:} lines in the tar file .......\\
 \emph{Card punching code:} ASCII\\
 \emph{Nature of physical problem:} A good starting description of
the electronic structure of extended many-electron systems such as
molecules, clusters, and polymers, can be obtained using the Hartree-Fock
(HF) method. Solution of HF equations within a fully \emph{ab initio}
formalism for large systems, however, is computationally quite expensive.
For such systems, semi-empirical methods such as CNDO and INDO proposed
by Pople and collaborators are quite attractive. The present program
can solve the HF equations for both open- and closed-shell systems
containing first- and second-row atoms using either the INDO model
or the CNDO model. \\
 \emph{Method of Solution:} The single-particle HF orbitals are expressed
as linear combinations of the Slater-type orbital (STO) basis set
specified by Pople and coworkers. Then using the parameters prescribed
for the CNDO/INDO methods, the HF integro-differential equations are
transformed into a matrix eigenvalue problem. Thereby, its solutions
are obtained in a self-consistent manner, using methods of computational
linear algebra.\\
 \emph{Unusual features of the program:} None

\section{Introduction}

The linear combination of atomic orbitals (LCAO) method is one of
the most common approaches for solving the Schr\"odinger equation
for many-electron systems such as atoms, molecules, clusters, and
solids. It consists of expressing the single-particle orbitals of
the electrons of the system as a linear combination of a known basis
set, and then solving the mean-field equations such as the Hartree-Fock
(HF) or the Kohn-Sham equations. This converts these integro-differential
equations into a matrix eigenvalue problem, which is subsequently
solved using computational approaches from the linear algebra. If
one intends to go beyond the mean-field to include the electron correlations
effects, approaches such as the configuration-interaction (CI), coupled-cluster,
or the Green's function based formalisms can be used. If $N$ is the
total number of basis functions used, the computational difficulty
at the mean-field level scales roughly as $N^{4}$, which is the number
of two-electron integrals needed to perform such calculations. For
post mean-field correlated calculations, integrals need to be transformed
from the basis-set atomic orbital (AO) representation to the molecular-orbital
representation (MO), a process which scales as $N^{5}$, while subsequent
solution of the corresponding equations can be even more time consuming\cite{jensen}.
Since $N$ increases rapidly with the number of atoms (and hence electrons)
in the system, therefore, for very large systems solution of even
the mean-field equations can become computationally intractable. Therefore,
it is always advisable to devise methods of electronic structure theory
which aim at reducing the size of the basis set.

Using the zero-differential overlap (ZDO) approximation developed
by Parr\cite{ZDO}, Pople and coworkers developed a series of semi-empirical
methods for computing the electronic structure of molecules such as
the Pariser-Parr-Pople (PPP) model\cite{PPP}, the complete neglect
of differential overlap (CNDO) method\cite{CNDO-1a,CNDO-1b,CNDO-2},
and the intermediate neglect of differential overlap (INDO) method\cite{INDO}.
Of these, the PPP model is applicable only to $\pi$-conjugated systems,
however, the CNDO and INDO models with suitable parametrization, are
in principle, applicable to all molecular systems\cite{pople-book}.
CNDO and INDO methods are a class of valence-electron models which
utilize a minimal Slater-type orbital (STO) basis set for the representation
of the valence orbitals\cite{pople-book}. Additionally, in the representation
of the Hamiltonian, only one- and two-center integrals are retained,
leading to a drastic reduction in the computational effort as compared
to the \emph{ab initio} calculations\cite{pople-book}. Therefore,
the CNDO/INDO models share attractive feature of semi-empirical parametrization
with the PPP model, and a spatial representation of the molecular
orbital with the \emph{ab initio} approaches\cite{pople-book}. And,
unlike the PPP model, the CNDO/INDO methods can also be used for the
geometry optimization of molecules\cite{pople-book}. Therefore, for
large molecular systems and clusters, for which the applications of
fully \emph{ab initio }approaches can be computationally intractable,
the CNDO/INDO methods provide an attractive alternative for the theoretical
description of their electronic structure.

It is with possible applications to large molecules, clusters, and
polymers in mind that we have developed the present computer program
which implements the CNDO-2/INDO methods as formulated originally
by Pople and coworkers\cite{pople-book}. As per the original formulation
by Pople and coworkers\cite{pople-book}, the INDO method can be used
for the molecules containing the first-row atoms, while the CNDO/2
method is applicable to those containing both the first-, and the
second-row, atoms. The fact that the code has been written in a modern
programming language, \emph{viz.}, Fortran 90, allows it to utilize
dynamic memory allocation, thereby freeing it from various array limits,
and resultant artificial restrictions on the size of the molecules.
Thus our program can be used on a given computer until all its available
memory is exhausted. The present computer program can perform restricted
Hartree-Fock (RHF) calculations on closed-shell systems, and unrestricted-Hartree-Fock
(UHF) calculations on open-shell systems. Additionally, it also allows
one to compute properties such as the molecular dipole moment, Mulliken
population analysis, and linear-optical absorption spectrum under
the electric-dipole approximation. Apart from describing the computer
program, we also present several of its applications which include
various small molecules, fullerene C$_{60}$, and polymeric chains
consisting of carbon atoms (C-chain), and alternating boron and nitrogen
atoms (BN-chain).

The remainder of the paper is organized as follows. In section \ref{sec:theory}
we briefly review the theory associated with the CNDO/INDO approaches.
Next, in section \ref{sec:program} we discuss the general structure
of our computer program, and also describe its constituent subroutines.
In section \ref{sec:install} we briefly describe how to install the
program on a given computer system, and to prepare the input files.
Results of various example calculations using our program are presented
and discussed in section \ref{sec:results}. Finally, in section \ref{sec:conclusions},
we present our conclusions, as well as discuss possible future directions.

\section{Theory}

\label{sec:theory}

In this section we briefly review the theory associated with the CNDO/INDO
methods. The detailed discussion on the topic can be found in the
book by Pople and Beveridge\cite{pople-book}. Our discussion will
be in the context of the UHF method, the corresponding RHF equations
can be easily deduced from them. As per the assumptions of the UHF
method, we assume that the $i$-th up- and down-spin orbitals are
distinct, and are represented (say) as $\psi_{i}^{(\alpha)}$ and
$\psi_{i}^{(\beta)}$, respectively. We assume that these orbitals
can be written as a linear combination of a finite-basis set \begin{equation}
\psi_{i}^{(\alpha)}=\sum_{\mu}C_{\mu i}^{(\alpha)}\phi_{\mu},\label{eq:lcao}\end{equation}
where $\phi_{\mu}$'s represent the basis functions in question, and
the determination of the unknown coefficients $C_{\mu i}^{(\alpha)}$
amounts to the solution of the UHF equations. In the equation above,
we have only stated the expressions for the up-spin orbitals, the
case of the down-spin orbitals can be easily deduced. Assuming the
Born-Oppenheimer Hamiltonian for the electrons of the system\begin{equation}
H=-\frac{\hbar^{2}}{2m}\sum_{i=1}^{N_{e}}\nabla_{i}^{2}-\sum_{A=1}^{N_{n}}\sum_{i=1}^{N_{e}}\frac{Z_{A}e^{2}}{R_{Ai}}+\sum_{i>j}\frac{e^{2}}{r_{ij}},\label{eq:born-opp}\end{equation}
 where the first term represents the kinetic energy of $N_{e}$ electrons
of the system, the second term represents the interaction energy of
those electrons with its $N_{n}$ nuclei, $Z_{A}$ represents the
nuclear charge of the $A$-th nucleus, $R_{Ai}$ denotes the distance
between that nucleus and the $i$-th electron, $r_{ij}$ represents
the inter-electronic distance, while $m$ and $e$ are electronic
mass and charge, respectively. We further assume that the total number
of up-/down-spin electrons is $N_{\alpha}/N_{\beta}$, such that $N_{\alpha}+N_{\beta}=N_{e}$.
Using the conjecture of Eq. \ref{eq:lcao} in conjunction with the
Hamiltonian above, one obtains the so-called Pople-Nesbet equations

\begin{equation}
\sum_{\nu}(F_{\mu\nu}^{\alpha}-\varepsilon_{i}^{\alpha}S_{\mu\nu})C_{vi}^{(\alpha)}=0,\label{eq:pople-nesbet}\end{equation}
where $S_{\mu\nu}$ is the basis function overlap matrix, $\epsilon_{i}^{\alpha}$
is the UHF eigenvalue of the $i$-th up-spin orbital, $F_{\mu\nu}^{\alpha}$
is the Fock matrix for the up-spin electrons defined by \begin{equation}
F_{\mu\nu}^{\alpha}=h_{\mu\nu}+\sum_{\lambda\sigma}[P_{\lambda\sigma}(\mu\nu|\lambda\sigma)-P_{\lambda\sigma}^{\alpha}(\mu\sigma|\lambda\nu)],\label{eq:fock}\end{equation}

above $h_{\mu\nu}$ represents the matrix elements of the one-electron
part (kinetic energy and the electron-nucleus interaction) of the
Hamiltonian of Eq. \ref{eq:born-opp}, $(\mu\nu|\lambda\sigma)$ represents
two-electron Coulomb repulsion integral in the Mulliken notation

\begin{equation}
(\mu\nu\mid\lambda\sigma)=\int\int d\tau_{1}d\tau_{2}\phi_{\mu}(1)\phi_{\nu}(1)r_{12}^{-1}\phi_{\lambda}(2)\phi_{\sigma}(2),\label{eq:two-el}\end{equation}
and $P_{\lambda\sigma}^{\alpha}$ and $P_{\lambda\sigma}$, are the
up-spin and total density matrix elements, respectively, defined as
\begin{equation}
P_{\lambda\sigma}^{\alpha}=\sum_{i=1}^{N_{\alpha}}C_{\lambda i}^{(\alpha)*}C_{\sigma i}^{(\alpha)},\label{eq:den-up}\end{equation}
and\begin{equation}
P_{\lambda\sigma}=P_{\lambda\sigma}^{\alpha}+P_{\lambda\sigma}^{\beta}.\label{eq:den-tot}\end{equation}

Equations \ref{eq:fock} through \ref{eq:den-tot}, define the UHF
method without any approximations. Next we briefly describe the approximations
involved in the CNDO/INDO methods, leading up to corresponding UHF
equations\cite{pople-book}:

\begin{enumerate}
\item Only valence electrons are treated explicitly thus $N_{e}=N_{v}$,
where $N_{v}$ represents the number of valence electrons in the system.
\item A STO basis set centered on the individual atoms of the system is
employed, with the basis functions of the form\begin{equation}
\phi_{\mu}(r,\theta,\phi)=R_{nl}^{\mu}(r)Y_{lm}(\theta,\phi),\label{eq:basis}\end{equation}
where $n,l,m$ represent the principal, orbital, and magnetic quantum
numbers associated with the $\mu$-th basis function, $Y_{lm}(\theta,\phi)$
is the real spherical harmonic, and the radial part of the basis function
is given by \begin{equation}
R_{nl}^{\mu}(r)=(2\zeta_{\mu})^{n+1/2}(2n!)^{-1/2}r^{n-1}\exp(-\zeta_{u}r),\label{eq:radial}\end{equation}
where $\zeta_{\mu}$ is the orbital exponent associated with the $\mu$-th
basis function, and is atom specific. In the CNDO method a minimal
basis set is employed for the first row atoms, while an augmented
basis set consisting also of $d$-type functions is employed for the
second-row atoms. The present implementation of the INDO method, which
is restricted only the to the first-row atoms, uses a basis set identical
to the CNDO method.
\item The effective one-electron matrix elements $h_{\mu\nu}$ are called
core integrals, and determined semi-empirically. Diagonal elements
of the one-electron element $h_{\mu\mu}$ are determined through various
parameters such as electron affinity $A_{\mu}$, and ionization potential
$I_{\mu}$ of the atoms involved, while the off-diagonal elements
($\mu\neq\nu)$ are determined by\begin{equation}
h_{\mu\nu}=\beta_{AB}^{0}S_{\mu\nu},\label{eq:core-int}\end{equation}
where $A$ and $B$ denote the atoms on which basis functions $\mu$
and $\nu$ are located, $\beta_{AB}^{0}$ is a semiempirical parameter
dependent on $A$ and $B$, and $S_{\mu\nu}$ is the overlap matrix
element for basis functions $\mu$ and $\nu$.
\item For orbital orthonormalization purposes it is assumed that the basis
set is orthonormal.
\item For the two-electron integrals $(\mu\nu\mid\lambda\sigma)$, following
approximation is adopted\begin{equation}
(\mu\nu\mid\lambda\sigma)=\delta_{\mu\nu}\delta_{\lambda\sigma}(\mu\mu\mid\lambda\lambda),\label{eq:two-int}\end{equation}
and this set of integrals is further reduced by assuming\begin{equation}
(\mu\mu\mid\lambda\lambda)=\gamma_{AB},\label{eq:two-int-gamma}\end{equation}
 where it is assumed that basis functions $\mu$ and $\lambda$ belong
to atoms $A$ and $B$, respectively. The value $\gamma_{AB}$ is
computed using the $s$-type orbitals located on $A$ and $B$ . Thus,
all the two-electron integrals, apart from one- and two-center integrals,
are ignored. As compared to the CNDO method, the following one-center
integrals of the type $(\mu\nu|\mu\nu)$ are assumed nonzero in the
INDO method. The values of these integrals are determined semiempirically
through the Slater-Condon parameters.
\end{enumerate}
Once all the approximations listed above are implemented, the diagonal
elements of the Fock matrix for the CNDO/2 model are given by\begin{equation}
F_{\mu\mu}^{\alpha}=-\frac{1}{2}(I_{\mu}+A_{\mu})+\sum_{B}(P_{BB}-Z_{B})\gamma_{AB}-(P_{\mu\mu}^{\alpha}-1/2)\gamma_{AA},\label{eq:fock-cndo-diag}\end{equation}
while the off-diagonal elements for both the CNDO-2 and the INDO are\begin{equation}
F_{\mu\nu}^{\alpha}=\beta_{AB}^{0}S_{\mu\nu}-P_{\mu\nu}^{\alpha}\gamma_{AB}.\label{eq:fock-cndo-off}\end{equation}
In the equations above, $Z_{B}$ represents the effective nuclear
charge of atom $B$, and $P_{BB}=\sum_{\mu\in B}P_{\mu\mu}$ is the
sum of those diagonal elements of the total density matrix which are
centered on atom $B$. In the INDO method, however, one uses different
expressions for the one-center diagonal and off-diagonal elements,
given by\begin{eqnarray}
F_{\mu\mu}^{\alpha} & = & U_{\mu\mu}+\sum_{\lambda\in A}[P_{\lambda\lambda}(\mu\mu|\lambda\lambda)-P_{\lambda\lambda}^{\alpha}(\mu\lambda|\mu\lambda)]\nonumber \\
 &  & +\sum_{B\neq A}(P_{BB}-Z_{B})\gamma_{AB},\label{eq:fock-indo-diag}\end{eqnarray}
 and \begin{equation}
F_{\mu\nu}^{\alpha}=(2P_{\mu\nu}-P_{\mu\nu}^{\alpha})(\mu\nu|\mu\nu)-P_{\mu\nu}^{\alpha}(\mu\mu|\nu\nu),\label{eq:fock-indo-off}\end{equation}
where $\mu,\nu\in A$. Above $U_{\mu\mu}$, and the one-center two
electron integrals are obtained through $I_{\mu}$, $A_{\mu}$, and
various Slater-Condon parameters\cite{INDO}. The two-center off-diagonal
elements of the Fock matrix for the INDO model are obtained through
\ref{eq:fock-cndo-off}. Once the Fock matrix is constructed, both
for the CNDO/2 and INDO models, one solves the eigenvalue problem
for the up-spin Fock matrix \begin{equation}
\sum_{\nu}F_{\mu\nu}^{\alpha}C_{\nu i}^{(\alpha)}=\varepsilon_{i}^{\alpha}C_{\mu i}^{(\alpha)},\label{eq:uhf-eigval}\end{equation}

as well as the down-spin Fock matrix, using the iterative diagonalization
technique, to achieve selfconsistency. From the equations given above,
it is easy to deduce the expressions for $F_{\mu\nu}^{\beta}$, as
well as the Fock matrix elements for the RHF case.

\section{Description of the Program}

\label{sec:program}

Our computer code consists of the main program, and various subroutines
and modules, all of which have been written in Fortran 90 language.
Additionally, the program must link to the LAPACK/BLAS library, whose
diagonalization routines are used by our program. In the following
we briefly describe the main program, and each subroutine.

\subsection{Main program CINDO}

This is the main program of our package which reads input data such
as atomic numbers of the atoms constituting the system, and their
positions, from the input file. The program also calculates the number
of valence electrons of the system under consideration, and the total
number of basis functions needed. It dynamically allocates various
arrays, and then calls other subroutines to accomplish the remainder
of the calculations. Because of the dynamical array allocation, the
user need not worry about various array sizes, as the program will
automatically terminate when it exhausts all the available memory
on the computer.

\subsection{Subroutine BASEGEN}

This subroutine generates various arrays containing information to
the basis functions used in the calculations. This includes quantities
such as principal quantum number ($n)$, orbital angular momentum
($l$), magnetic quantum number ($m$), orbital exponent ($\zeta_{\mu}$)
associated with each STO type basis function defined in Eqs. \ref{eq:basis}
and \ref{eq:radial}. Additionally, it also stores some semi-empirical
data associated with the Hamiltonian such the $\beta_{\mu\nu}^{0}$,
and various Slater-Condon parameters. This routine is called from
the main program CINDO.\texttt{ }

\subsection{Subroutine FACTCAL}

The primary task of this subroutine is to generate the factorials
of various integers. The factorials thus generated are stored in global
arrays accessible via the MODULE \texttt{factorials}. This subroutine
is also called from the main program CINDO.

\subsection{Subroutine ASSOC\_LEGNDRE }

This subroutine initializes the expansion coefficients which define
associated Legendre polynomials of various degrees, needed to represent
the angular part of the basis functions. The data is stored in global
arrays through MODULE \texttt{legendre.} This subroutine is also called
from the main program.

\subsection{function SS }

A very important quantity used in computing Hamiltonian matrix elements
is the so-called reduced overlap integral between two basis functions
(labeled $a$, and $b$)\cite{pople-book}\begin{eqnarray}
s(n_{a},l_{a},m,n_{b},l_{b},\alpha,\beta) & = & \int_{1}^{\infty}\int_{-1}^{1}(\mu+\nu)^{n_{a}}(\mu-\nu)^{n_{b}}\exp[-\frac{1}{2}(\alpha+\beta)\mu]\nonumber \\
 &  & \times\exp[-\frac{1}{2}(\alpha-\beta)\nu]T(\mu,\nu)d\mu d\nu,\label{eq:redovp}\end{eqnarray}
where\begin{eqnarray}
T(\mu,\nu) & = & D(l_{a},l_{b},m)\sum_{u}^{l_{a}-m}\sum_{v}^{l_{b}-m}C_{l_{a}mu}C_{l_{b}mu}(\mu^{2}-1)^{m}(1-\nu^{2})^{m}\nonumber \\
 &  & \times(1+\mu\nu)^{u}(1-\mu\nu)^{v}(\mu+\nu)^{-m-u}(\mu-\nu)^{-m-v}.\label{eq:tmunu}\end{eqnarray}

Above $(n_{a},l_{a},m)$ and $(n_{b},l_{b,}m)$ are the quantum numbers
of two basis functions, $C_{l_{a}mu}$, $D(l_{a},l_{b},m)$ etc. are
coefficients associated with the angular part of the basis functions,
and $\alpha=\zeta_{a}R$, and $\beta=\zeta_{b}R$, where $\zeta_{a}$,
$\zeta_{b}$ are the basis function exponents, and $R$ is the distance
between the atoms on which basis functions are located. If we define
the so-called $Y_{ij\lambda}$ coefficients defined through the relation\begin{eqnarray}
\sum_{u}^{l_{a}-m}\sum_{v}^{l_{b}-m}C_{l_{a}mu}C_{l_{b}mu}(\mu^{2}-1)^{m}(1-\nu^{2})^{m}\nonumber \\
\times(1+\mu\nu)^{u}(1-\mu\nu)^{v}(\mu+\nu)^{n_{a}-m-u}(\mu-\nu)^{n_{b}-m-v} & = & \sum_{i,j=0}Y_{ij\lambda}\mu^{i}\nu^{j},\label{eq:ycoef}\end{eqnarray}
we obtain the expression\begin{equation}
s(n_{a},l_{a},m,n_{b},l_{b},\alpha,\beta)=D(l_{a},l_{b},m)\sum_{i,j}Y_{ij\lambda}A_{i}[\frac{1}{2}(\alpha+\beta)]B_{j}[\frac{1}{2}(\alpha-\beta)],\label{eq:sint}\end{equation}
where \begin{equation}
A_{k}(\rho)=\int_{1}^{\infty}x^{k}\exp(-\rho x)dx,\label{eq:aint}\end{equation}
and \begin{equation}
B_{k}(\rho)=\int_{-1}^{1}x^{k}\exp(-\rho x)dx.\label{eq:bint}\end{equation}

For the $s$ functions ( $l_{a}=l_{b}=m=0$), the reduced overlap
integrals (cf. Eq. \ref{eq:redovp}) can be written as\begin{eqnarray}
s(n_{a},0,0,n_{b},0,\alpha,\beta) & = & \frac{1}{2}\int_{1}^{\infty}\int_{-1}^{1}(\mu+\nu)^{n_{a}}(\mu-\nu)^{n_{b}}\exp[-\frac{1}{2}(\alpha+\beta)\mu]\nonumber \\
 &  & \times\exp[-\frac{1}{2}(\alpha-\beta)\nu]d\mu d\nu.\label{eq:redovp-s}\end{eqnarray}
If we define the so-called $Z_{k\lambda}$ coefficients through\begin{equation}
(\mu+\nu)^{n_{a}}(\mu-\nu)^{n_{b}}=\sum_{k=0}^{n_{a}+n_{b}}Z_{k\lambda}\mu^{k}\nu^{(n_{a}+n_{b}-k)},\label{eq:zcoef}\end{equation}
we obtain\begin{equation}
s(n_{a},0,0,n_{b},0,\alpha,\beta)=\frac{1}{2}\sum_{k=0}^{n_{a}+n_{b}}Z_{k\lambda}A_{k}[\frac{1}{2}(\alpha+\beta)]B_{n_{a}+n_{b}-k}[\frac{1}{2}(\alpha-\beta)].\label{eq:sint-s}\end{equation}

The task of this REAL(kind=8) function is to compute the reduced overlap
integral as defined in Eqs. \ref{eq:sint} and \ref{eq:sint-s}, for
a given pair of basis functions $a$ and $b$. The input to this routine
is all the basis function related information such as their quantum
numbers, orbital exponents, and the distance between them. It performs
these calculations by calling subroutines GETYCOEF, GETZCOEF, AINT,
and BINT which described below.

\subsection{Subroutine GETYCOEF}

The task of this subroutine is to compute these $Y_{ij\lambda}$ coefficients,
for a given set of $n_{a}$, $n_{b}$, $l_{a}$, $l_{b}$, and $m$
as defined in Eq. \ref{eq:ycoef}. It achieves this goal by calling
subroutines BINOMIAL and POL2MUL described below.

\subsection{Subroutine GETZCOEF }

The task of this subroutine is to compute the $Z_{k\lambda}$ coefficients,
defined in Eq. \ref{eq:zcoef}, for a given pair of $s$-type basis
functions. As in case of subroutine GETYCOEF, this routine also computes
for these coefficients by calling routines BINOMIAL and POL2MUL.

\subsection{Subroutine BINOMIAL}

Using the binomial expansion, expression $(ax^{m}y^{n}+bx^{p}y^{q})^{l}$
can be expanded as\begin{equation}
(ax^{m}y^{n}+bx^{p}y^{q})^{l}=\sum_{i,j}c_{ij}x^{i}y^{j},\label{eq:binom}\end{equation}
 where $i$, $j$, $m$, $n$, $p$, $q$, and $l$, are integers,
$x$, and $y$ are variables, and $a$, $b$, and $c_{ij}$'s are
constants. This subroutine computes these expansion coefficients $c_{ij}$'s
for a given set of input values of $a$, $b$, $m$, $n$, $p$, $q$,
and $l$. It is called both from routines GETZCOEF and GETYCOEF.

\subsection{Subroutine POL2MUL }

This subroutine computes the coefficients of the product polynomial
when two polynomial of the type $\sum_{i,j}a_{ij}x^{i}y^{j}$are multiplied,
\emph{i.e.}, \begin{equation}
\sum_{i,j}c_{ij}x^{i}y^{j}=(\sum_{k,l}a_{kl}x^{k}y^{l})(\sum_{l,m}b_{lm}x^{l}y^{m}).\label{eq:pol-prod}\end{equation}

The input to this routine are coefficients $a_{kl}$ and $b_{kl}$,
while the output consists of $c_{ij}$. The arrays meant for storing
these coefficients are allocated dynamically.

\subsection{Subroutine AINT }

The value of the integral $A_{k}(\rho)$, defined in Eq. \ref{eq:aint},
can be shown to be\begin{equation}
A_{k}(\rho)=\exp(-\rho)\sum_{\mu=1}^{k+1}\frac{k!}{\rho^{\mu}(k-\mu+1)!}.\label{eq:a-series}\end{equation}
 Subroutine AINT uses this series to compute the value of $A_{k}(\rho)$,
for a given value of $k$ and $\rho$.

\subsection{Subroutine BINT }

The purpose of this subroutine is to compute integral $B_{k}(\rho)$,
defined in Eq. \ref{eq:bint}, for a given value of $k$ and $\rho$.
We use the following recursion relation to perform the task\begin{eqnarray}
B_{k+1}(\rho) & = & -A_{k+1}(\rho)+\frac{(-1)^{^{k+1}}\exp(\rho)}{\rho}\nonumber \\
 &  & +(k+1)\left(\frac{(A_{k}(\rho)+B_{k}(\rho))}{\rho}\right).\label{eq:bformula}\end{eqnarray}
Thus first a call is made to the routine AINT to compute all the $A_{k}(\rho)$
's needed. Subsequently, the $B_{k}(\rho)$'s are generated using
the recursion relation of Eq. \ref{eq:bformula}.

\subsection{Subroutine REDOVINT}

This is a very important subroutine which evaluates overlap matrix
elements $S_{\mu\nu}$ among the basis functions. It evaluates the
reduced overlap integrals described above for each pair of basis functions,
by calling the function SS, using a coordinate system in which the
atoms corresponding to the basis function pair are located along the
$z$-axis. Then by a call to the subroutine TRANS described below,
it obtains the actual overlap integrals by transforming the reduced
integrals from the special coordinate system, to the actual molecular
coordinate system. The upper-triangle of the overlap matrix is stored
in a one-dimensional array.

\subsection{Subroutine TRANS xmgrace}

The formulas for reduced overlap integrals (Eqs. \ref{eq:redovp}
and \ref{eq:redovp-s}) assume that the atoms on which the basis functions
are centered are a distance $R$ apart from each other along the $z$-axis.
But in practice, the molecules may have any kind of orientation. Therefore,
we need to transform the reduced overlap integrals computed using
these formulas, to the real orientation of the molecule. This is achieved
through a transformation matrix which depends upon angular momenta
of the basis functions, as well as on the angles by which the $z$-axis
should be rotated to align it with the real orientation of the atoms
involving the two basis functions. The task of this subroutine is
to construct this transformation matrix, and then apply it to obtain
the overlap integrals with respect to the molecular frame.

\subsection{Subroutine COUL\_INT }

This subroutine calculates the Coulomb integrals $\gamma_{AB}$ (\emph{cf.}
Eq. \ref{eq:two-int-gamma}) needed for the construction of the Fock
matrix. It can be shown that these integrals are proportional to the
reduced overlap integrals discussed above. Therefore, this routine
computes these integrals by calling the function SS, and stores the
values (one per atom pair) in a two-dimensional array.

\subsection{Subroutine CORE\_INT }

The aim of this subroutine is to compute the one-electron part of
Fock matrix, referred to as core integrals, and discussed in section
\ref{sec:theory}. The calculation of off-diagonal elements involves
the use of the overlap matrix elements $S_{\mu\nu}$ computed in the
routine REDOVINT, discussed earlier. The semiempirical data needed
for computing these matrix elements is also passed to this routine
through arguments. The upper-triangle of the one-electron part of
the Fock matrix, along with the extended H\"uckel Hamiltonian, are
finally stored in separate one-dimension arrays, and constitute the
output of this routine.

\subsection{Subroutine DIPINT }

The aim of this subroutine is to compute matrix elements of dipole
operator over the basis set. This subroutine is called only if the
linear-optical absorption, or permanent electric dipole calculations
are desired. Standard formulas are utilized to compute these matrix
elements, and it is called from the main program CINDO.

\subsection{Subroutine SCF\_RHF }

This subroutine solves the RHF equations for the system under consideration
in a self-consistent manner, using the iterative diagonalization procedure.
The arrays which are needed during the calculations are allocated
before the calculations begins, and are deallocated upon completion.
Before the first iteration, extended H\"uckel Hamiltonian is diagonalized
to obtain a set of starting orbitals. Subsequently, the Fock matrix
corresponding to those orbitals is constructed and diagonalized. The
process is repeated until the self-consistency is achieved. During
the self-consistency iterations, subroutine DSPEVX from the LAPACK/BLAS
library is used to obtain the occupied eigenvalues and eigenvectors.
Obtaining only the occupied eigenpairs, as against the entire spectrum,
leads to considerable savings of CPU time for large systems. However,
if the entire spectrum of eigenvalues and eigenvectors is needed,
say, to perform optical absorption calculations, the Fock matrix is
diagonalized using the routine DSPEV from the LAPACK/BLAS library,
upon completion of the self-consistency iterations. Because the entire
spectrum is obtained only after self-consistency has been achieved,
it does not strain the computational resources too much. Apart from
computing the RHF total energy, this subroutine also calculates the
total binding energy of the system, and, if needed, performs Mulliken
population analysis as well.

\subsection{Subroutine SCF\_UHF }

This subroutine is exactly the same in its logic and structure as
the previously described SCF\_RHF, except that the task of this routine
is to solve the UHF equation for the system under consideration. Different
Fock matrices for the up- and the down-spin are constructed and diagonalized
in each iteration, until the self-consistency is achieved. Similar
to the case of routine SCF\_RHF, during the iterations only the occupied
eigenvalues and eigenvectors are computed using the routine DSPEVX.
The iterations are stopped once the total UHF energy of the system
converges to within a user defined threshold.

\subsection{Subroutine PROPERTY }

This is a driver subroutine whose task is to read the converged SCF
orbitals written onto the disk by the SCF routines, and then call
other subroutines meant for computing various properties of the system
under investigation. It is called from the main program CINDO after
the SCF calculations, provided the user has opted for one of the property
calculations such as the permanent electric dipole moment of the molecule,
or its optical absorption spectrum.

\subsection{Subroutine DIPIND }

This subroutine transforms the dipole matrix elements from the basis-set
AO representation to the SCF MO representation, by means of a two-index
transformation. Therefore, it uses the dipole matrix elements computed
in DIPINT, and the SCF MOs as inputs. The transformed dipole matrix
elements, which constitute the output of this routine, are used in
the calculation of linear optical absorption spectrum of the molecule.
This subroutine is called from the routine PROPERTY described above,
if the user has opted for the optical absorption calculations.

\subsection{Subroutine DIPMOM\_RHF }

This subroutine calculates the total net electric dipole moment component
of the molecule under investigation for restricted Hartree-Fock case.
It is called from the routine PROPERTY if the user has opted for the
dipole moment calculation. It uses dipole matrix elements calculated
in the subroutine DIPINT and the SCF MOs as input, and computes the
permanent dipole moment of the system using a straightforward formula.

\subsection{Subroutine DIPMOM\_UHF }

The purpose and logic of this routine is the same as DIPMOM\_RHF,
except that it is used for the case when UHF calculations have been
performed. This routine is also called from the subroutine PROPERTY.

\subsection{Subroutine SPECTRUM}

This is an important subroutine which calculates the linear optical
absorption of the system, under electric-dipole approximation, assuming
a Lorentzian line shape and a constant line width for all the levels.
Thus, if this calculation is opted, in the input file the user needs
to provide the line width, along with the range of frequencies over
which the spectrum needs to be computed. Additionally, the routine
uses the dipole matrix elements over the MOs as computed in routine
DIPIND, along with the RHF single-particle energies. The computed
spectrum is written in an ASCII file 'spectrum.dat', which can be
readily used for plotting using programs such as gnuplot\cite{gnuplot}
or xmgrace\cite{xmgrace}. This subroutine is also called from the
routine PROPERTY, if the user has opted for linear absorption spectrum
calculations.

\subsection{Orbital and Charge Density Plotting Subroutines}

For the purpose of orbital visualization, our code offers several
options to the user for plotting the MOs, and the corresponding charge
density. It is accomplished through four subroutines, PLOT\_1D\_RHF,
PLOT\_2D\_RHF, PLOT\_1D\_UHF, and PLOT\_2D\_UHF.

The task of subroutine PLOT\_1D\_RHF is to compute and print out the
numerical values of RHF MOs, or their charge densities, on a one-dimensional
grid of points, whose direction and range is provided by the user.
Output of this program is written in an ASCII file called 'orbplot.dat',
and can be readily used for plotting by gnuplot\cite{gnuplot} and
xmgrace\cite{xmgrace}. This routine is called from the routine PROPERTY,
if the user has opted for it. To compute the numerical values of RHF
MOs (or their charge densities) at different points in space, it uses
the numerical values of basis functions computed at those points,
by calling function BASFUNC.

When a user is interested in obtaining a two-dimensional plot of the
RHF orbitals/charge densities in the Cartesian planes, subroutine
PLOT\_2D\_RHF\texttt{ }is called from the routine PROPERTY. The structure
of this routine is also similar to that of PLOT\_1D\_RHF, except that
for this case the orbital/density values are printed out with respect
to the two cartesian coordinates of the plane. This routine also uses
function BASFUNC to compute the numerical values of the orbitals/densities,
and the output is also written in the file 'orbplot.dat'. In order
to facilitate contour plots of charge densities, the option of making
logarithmic plots is also available.

In case of open-shell UHF calculations, the corresponding plots of
the up- and down-spin MOs are obtained through calls to subroutines
PLOT\_1D\_UHF and PLOT\_2D\_UHF, and the output is again written in
the file 'orbplot.dat'.

\subsection{Function BASFUNC }

It is a REAL(kind=8) function whose aim is to calculate the numerical
value of a given basis function, at a particular point in space. Therefore,
the input to this function consists of the coordinates of the point
in space with respect to the location of the basis function, $(n,l,m)$
quantum numbers of the basis function, and its exponent $\zeta$.
This function is called from all the orbital/density plotting subroutines
described above.

\section{Installation, input files, output files}

\label{sec:install}

We believe that the installation and execution of the program, as
well as preparation of suitable input files is fairly straightforward.
Therefore, we will not discuss these topics in detail here. Instead,
we refer the reader to the README file for details related to the
installation and execution of the program. Additionally, the file
'\texttt{input\_prep.pdf}' explains how to prepare a sample input
file. Several sample input and output files corresponding to various
example runs are also provided with the package.

\section{Results and Discussions}

\label{sec:results}

In this section, we present and discuss the numerical applications
of our results. First we present the results on a number of molecules.
Next, we apply our method to obtain the ground states of model polymeric
systems C chain and BN chain. Finally, we present the results of our
calculations of optical absorption in Buckminster fullerene C$_{60}$.
Wherever possible, we compare our results to those published by other
authors.

\subsection{Molecular Systems}

\label{sub:molecules}

In this section we present the results of our calculations on a variety
of molecules, including fullerene C$_{60}$. The aim of these calculations
is to compare our results with those published by other authors\cite{pople-book},
and also with the CNDO/INDO calculations performed using Gaussian
03\cite{gaussian03}, in order to check the correctness of our program. 

In table \ref{tab:ehf} we compare the total HF energies of several
 molecules computed by our program, to those computed using Gaussian
03\cite{gaussian03}. We used the bond length of $0.74$ \AA\  for
the H$_{2}$ molecule as used also by Surjan\cite{surjan}. For water
molecule we used the geometry from Schaeffer \emph{et al}. \cite{h2o-geom},
for formic acid from Schwartz \emph{et al}.\cite{formic-geom}, for
borazane from Palke\cite{borazane-geom}, and for fluoropropene from
the work of Scarzafava \emph{et al}.\cite{propene-geom}. For C$_{60}$,
we considered a dimerized configuration with the $I_{h}$ symmetry
group and bond lengths 1.449 \AA\  and 1.397 \AA\  optimized by
Shibuya and Yoshitani \cite{shibu-yoshi}. The Cartesian coordinates
for the carbon atoms of C$_{60}$ were generated using the computer
program developed by Dharamvir and Jindal\cite{c60-coord}. Thus,
for all the cases illustrated in the table, the agreement on the total
HF energies between our calculations and those obtained using Gaussian
03\cite{gaussian03} is excellent both for the CNDO/2 and the INDO
methods. 

\begin{table}
\caption{Comparison of the total Hartree-Fock energies ($E_{HF}$) of several
molecules obtained using both the CNDO/2 and the INDO methods using
our program, to those computed using Gaussian 03\cite{gaussian03}.
All results are in atomic units. S/E inside the parentheses imply
staggered/eclipsed configurations. For the molecular geometries utilized
in these calculations, refer to the text.}

\begin{tabular}{|c|c|c|c|c|}
\hline 
Molecule & \multicolumn{2}{c|}{$E_{HF}$(This work)} & \multicolumn{2}{c|}{$E_{HF}$(Gaussian03)}\tabularnewline
\hline
 & CNDO/2 & INDO & CNDO/2 & INDO\tabularnewline
\hline
\hline 
H$_{2}$ & -1.474625 & -1.474625 & -1.474625 & -1.474625\tabularnewline
\hline 
H$_{2}$O & -19.868052 & -19.013606 & -19.868052 & -19.013606\tabularnewline
\hline
\emph{cis}-HCOOH & -45.305164 & -43.364618 & -45.305163 & -43.364618\tabularnewline
\hline 
\emph{trans}-HCOOH & -45.301984 & -43.360996 & -45.301984 & -43.360996\tabularnewline
\hline 
Borazane (E) & -20.169898 & -19.567825 & -20.169897 & -19.567824\tabularnewline
\hline 
Borazane (S) & -20.172764 & -19.570726 & -20.172763 & -19.570730\tabularnewline
\hline 
\emph{cis}-fluoropropene (E) & -52.763845 & -50.693901 & -52.763845 & -50.693901\tabularnewline
\hline
\emph{cis}-fluoropropene (S) & -52.762138 & -50.692238 & -52.762138 & -50.692238\tabularnewline
\hline
\emph{trans}-fluoropropene (E) & -52.761824 & -50.692176 & -52.761823 & -50.692175\tabularnewline
\hline
\emph{trans}-fluoropropene (S) & -52.759748 & -50.690019 & -52.759748 & -50.690019\tabularnewline
\hline
C$_{60}$ & -427.624631 & -412.293447 & -427.624631 & -412.293447\tabularnewline
\hline
\end{tabular}\label{tab:ehf}
\end{table}

Next we turn our attention to the comparison of results for geometry
optimization of a few closed- and open-shell molecules. In table \ref{tab:mol-geom}
we compare the bond lengths optimized by our program to those reported
by Pople \emph{et al.}\cite{pople-book} for several closed- and open-shell
diatomic molecules. Again the agreement obtained between the two sets
of calculations is excellent both for the CNDO/2 and the INDO calculations. 

\begin{table}
\caption{Comparison of geometries optimized by our code to those reported by
Pople \emph{et al}.\cite{pople-book}, for several small molecules.
Calculations for all the molecules with doublet or triplet ground
states were performed using the UHF method.}

\begin{tabular}{|c|c|c|c|c|c|}
\hline 
Molecule & State & \multicolumn{4}{c|}{Equilibrium Length (\AA)}\tabularnewline
\hline
\cline{3-4} \cline{5-6} 
 &  & \multicolumn{2}{c|}{This work} & \multicolumn{2}{c|}{Pople \emph{et al}.\cite{pople-book}}\tabularnewline
\cline{3-4} \cline{5-6} 
\hline 
 &  & CNDO/2 & INDO & CNDO/2 & INDO\tabularnewline
\hline
\hline 
Li$_{2}$ & $^{1}\Sigma_{g}^{+}$ & $2.179$ & $2.134$ & $2.179$ & $2.134$\tabularnewline
\hline 
B$_{2}$ & $^{3}\Pi_{g}$ & $1.278$ & $1.278$ & $1.278$ & $1.278$\tabularnewline
\hline 
C$_{2}$ & $^{1}\Sigma_{g}^{+}$ & $1.146$ & $1.148$ & $1.146$ & $1.148$\tabularnewline
\hline 
N$_{2}^{+}$ & $^{2}\Sigma_{g}^{+}$ & $1.127$ & $1.129$ & $1.127$ & $1.129$\tabularnewline
\hline
N$_{2}$ & $^{1}\Sigma_{g}^{+}$ & $1.140$ & $1.147$ & $1.140$ & $1.147$\tabularnewline
\hline
O$_{2}^{+}$ & $^{2}\Pi_{g}$ & $1.095$ & $1.100$ & $1.095$ & $1.100$\tabularnewline
\hline
O$_{2}$ & $^{3}\Sigma_{g}^{-}$ & $1.132$ & $1.140$ & $1.132$ & $1.140$\tabularnewline
\hline
NH & $^{3}\Sigma^{-}$ & $1.061$ & $1.069$ & $1.061$ & $1.070$\tabularnewline
\hline
OH & $^{2}\Pi_{i}$ & $1.026$ & $1.033$ & $1.026$ & $1.033$\tabularnewline
\hline
BeH & $^{2}\Sigma_{}^{+}$ & $1.324$ & $1.324$ & $1.324$ & $1.323$\tabularnewline
\hline
LiH & $^{1}\Sigma^{+}$ & $1.573$ & $1.572$ & $1.573$ & $1.572$\tabularnewline
\hline
BN & $^{3}\Sigma^{+}$ & $1.269$ & $1.269$ & $1.268$ & $1.269$\tabularnewline
\hline
LiF & $^{1}\Sigma^{+}$ & $2.161$ & $2.162$ & $2.161$ & $2.162$\tabularnewline
\hline
HF & $^{1}\Sigma^{+}$ & $1.000$ & $1.005$ & $1.000$ & $1.006$\tabularnewline
\hline
BF & $^{1}\Sigma_{}^{+}$ & $1.404$ & $1.408$ & $1.404$ & $1.408$\tabularnewline
\hline
\end{tabular}\label{tab:mol-geom}
\end{table}

Finally, in table \ref{tab:mol-edm} we compare the molecular dipole
moments and Mulliken populations of several heteronuclear diatomic
molecules obtained by our code with those reported by Pople \emph{et
al.} \cite{pople-book}. Both for the CNDO/2 and the INDO calculations
the agreement between our results and those of Pople \emph{et al.}
\cite{pople-book} is virtually exact. Thus, excellent agreement between
our results with those of other authors, not just for HF total energy,
but also for other properties, testifies to the essential correctness
of our computer program.

\begin{table}
\caption{Comparison of computed electric dipole moments and Mulliken populations
of heteronuclear diatomics molecules with the work of Pople \emph{et
al}.\cite{pople-book} The first number in each category is the CNDO/2
result, while the second number represents the INDO result.}

\begin{tabular}{|c|c|c|c|c|}
\hline 
Molecule & \multicolumn{2}{c|}{Electric Dipole Moment (Debye)} & \multicolumn{2}{c|}{Mulliken Population}\tabularnewline
\hline
\hline 
 & This work & Pople \emph{et al}.\cite{pople-book} & This work & Pople \emph{et al}.\cite{pople-book}\tabularnewline
\hline 
NH & 1.76/1.69 & 1.76/1.68 & 0.08/0.09 & 0.08/0.09\tabularnewline
\hline 
OH & 1.78/1.80 & 1.78/1.79 & 0.16/0.18 & 0.17/0.18\tabularnewline
\hline 
BeH & 0.67/0.65 & 0.67/0.64 & 0.14/0.14 & 0.14/0.14\tabularnewline
\hline 
LiH & 6.16/6.20 & 6.16/6.20 & 0.27/0.29 & 0.27/0.29\tabularnewline
\hline 
BN & 0.36/0.50 & 0.36/0.50 & 0.05/0.03 & 0.05/0.03\tabularnewline
\hline 
LiF & 7.91/7.87 & 7.90/7.86 & 0.56/0.58 & 0.56/0.58\tabularnewline
\hline
HF & 1.86/1.99 & 1.86/1.98 & 0.23/0.27 & 0.23/0.27\tabularnewline
\hline
BF & 1.31/0.87 & 1.31/0.86 & 0.15/0.15 & 0.15/0.15\tabularnewline
\hline
\end{tabular}\label{tab:mol-edm}

\end{table}

\subsection{Calculations on Lithium Clusters}

In this section we discuss the optimized geometries of small lithium
clusters computed using our program. The number of computational studies
of the electronic structure of small lithium clusters by other authors
is too numerous to list here. We will mainly refer to the \emph{ab
initio} works of Ray \emph{et al.}\cite{akray}, Boustani \emph{et
al.}\cite{boustani}, Jones \emph{et al}.\cite{hutter} \emph{,} and
Wheeler \emph{et al}.\cite{schaefer} who studied clusters similar
to the ones studied by us. Detailed computational studies of several
large atomic clusters containing various atoms are in progress in
our group, and will be published later.

\subsubsection{Li$_{2}$}

Results of our calculations on the optimized geometry of lithium dimer
for the closed-shell ground state were presented in Table \ref{tab:mol-geom}.
As is obvious from the table that our optimized bond lengths of 2.179
\AA (CNDO) and 2.134 \AA (INDO) for Li$_{2}$ are in perfect agreement
with similar calculations performed by Pople \emph{et al.}\cite{pople-book}.
As far as the comparison with the experiments is concerned, both these
results are significantly smaller than the measured value of 2.672
\AA\cite{herzberg}. Therefore, it will be of considerable interest
whether, or not, the inclusion of electron correlation effects will
improve the results.

\subsubsection{Li$_{3}$}

Geometrical configurations for a triatomic cluster can be broadly
classified as: (a) linear, and (b) triangular. For homonuclear systems
such as Li$_{3}$, the possible triangular geometries can be further
subclassified into: (i) equilateral triangle, (ii) isosceles triangle,
and (iii) a triangle with unequal arms. Of course, the equilateral
triangle geometry ($D_{3h}$) is expected to undergo Jahn-Teller distortion
to a lower symmetry configuration. Indeed, several density-functional
theory (DFT) and \emph{ab initio} correlated calculations have indicated
that the isosceles triangle geometry ($C_{2v}$) is the most stable
configuration for Li$_{3}$\cite{akray,boustani,hutter,schaefer}.
Our calculations were performed on the doublet ground state using
the UHF method, and the results are summarized in table \ref{tab:li3}.
We found that equilateral triangular configuration is energetically
more favorable as compared to the Jahn-Teller distorted isosceles
triangles, as well as equidistant linear configuration, both for CNDO
and INDO models. Optimized INDO and CNDO geometries are in very good
agreement with each other. The potential energy surface of the triangular
configuration shows interesting features for both the models. We find
that if the equal arms of the triangle are longer or shorter than
the optimized bond lengths of the $D_{3h}$ geometry presented in
table \ref{tab:li3}, the system does exhibit Jahn-Teller instability.
For bond lengths longer than those of the $D_{3h}$ geometry, the
distorted triangle has an angle less than $60^{o}$ between the equal
arms, while for bond lengths smaller than the optimized values, the
corresponding angle is more than $60^{o}$. However, the global minimum
was found for the $D_{3h}$ geometry described in table \ref{tab:li3}.
As far as the \emph{ab initio} correlated and the DFT calculations
are concerned, most of them report the length of equal arms of the
$C_{2v}$ geometry close to $2.8$ \AA, and the angle between them
in excess of $70^{o}$\cite{akray,boustani,hutter,schaefer}. Therefore,
it will be interesting whether the inclusion of electron correlation
effects will improve the agreement between CNDO/INDO models and the
\emph{ab initio} results.

\begin{table}
\begin{centering}
\begin{tabular}{|c|c|c|c|c|}
\hline 
Structure & \multicolumn{2}{c|}{CNDO Results} & \multicolumn{2}{c|}{INDO Results}\tabularnewline
\hline
\hline 
 & Bond Length (\AA) & $E_{HF}$(a.u.) & Bond Length (\AA) & $E{}_{HF}$(a.u.)\tabularnewline
\hline
\hline 
Linear & 1.461 & -1.8870412 & 1.457 & -1.8819986\tabularnewline
\hline
$D_{3h}$ & 1.932 & -2.0133753 & 1.919 & -2.0066737\tabularnewline
\hline
\end{tabular}
\par\end{centering}

\caption{Optimized CNDO and INDO geometries of Li$_{3}$, and corresponding
HF energies ($E_{HF}$). Calculations were performed on the doublet
ground states using the UHF method.}
\label{tab:li3}
\end{table}

\subsubsection{Li$_{4}$ }

For Li$_{4}$ clusters various geometries, ranging from linear to
tetrahedral are possible as investigated, \emph{e.g.}, by Ray \emph{et
al.}\cite{akray}. However, as reported by various authors, a rhombus
structure is energetically most favorable. As a demonstration of our
code we compute the relative stability of three possible structures
of this system namely, linear, square, and rhombus. 

\begin{table}
\begin{tabular}{|c|c|c|c|c|}
\hline 
Structure & \multicolumn{2}{c|}{CNDO Results} & \multicolumn{2}{c|}{INDO Results}\tabularnewline
\hline
\hline 
 & Bond Length (\AA) & $E_{HF}$(a.u.) & Bond Length (\AA) & $E_{HF}$(a.u.)\tabularnewline
\hline
\hline 
Linear  & 1.186 & -2.9683366 & 1.185 & -2.9590571\tabularnewline
\hline 
Square & 1.617 & -3.3083610 & 1.612 & -3.2976927\tabularnewline
\hline
\end{tabular}

\caption{Optimized geometries of Li$_{4}$ clusters of various shapes obtained
by CNDO and INDO methods, and corresponding HF energies ($E_{HF}$).
Calculations were performed on the closed-shell ground state.}

\label{tab:li4}
\end{table}

Results of our calculations are summarized in table \ref{tab:li4}.
We found that, of the three possible structures considered, the square
structure had the minimum energy. The rhombus shaped structures have
lower energy than the square structure for bond lengths in excess
of 2.1 \AA. However, the energies of those structures was found to
be higher than those of square structures reported in table \ref{tab:li4}.
This result is similar to what we obtained for Li$_{3}$ for which
the equilateral triangular structure was found to be more stable than
the isosceles triangular structure, both within the CNDO and INDO
models. Our result for Li$_{4}$ disagrees with those obtained by
correlated \emph{ab initio}, and DFT calculations\cite{akray,boustani,hutter,schaefer}
which predict the lowest energy for the rhombus structure with its
acute angle close to 50$^{o}$. Additionally, all \emph{ab initio}
calculations predict bond lengths significantly larger than obtained
here. Therefore, it is of considerable interest to explore whether
the inclusion of electron-correlation effects will bring our results
in better agreement with the \emph{ab initio} ones.

\subsection{Ground state of polymers}

Our code can be used to study both the ground and excited state properties
of oligomers of various polymers because they are nothing but finite
molecules, ranging in size from small to large. However, in this section
we demonstrate that our code can also be used to obtain the ground
state energy/cell, in the bulk limit, for one-dimensional periodic
systems such as polymers. Thus, it can be used, \emph{e.g.}, for the
purpose of ground-state geometry optimization of polymers, which is
what we demonstrate next.

The energy per unit cell of a one-dimensional periodic system can
be obtained using the formula\begin{equation}
E_{cell}=\lim_{n\rightarrow\infty}\Delta E(n)=\lim_{n\rightarrow\infty}(E(n+1)-E(n)),\label{eq:ecell}\end{equation}
where $E(n+1)/E(n)$ represent the total energies of oligomers containing
$n+1/n$ unit cells. Thus, using this formula, for sufficiently large
value of $n$, one can obtain the energy/cell of a polymer in the
bulk limit, from oligomer based calculations. In what follows we show
that value of $E_{cell}$ converges quite rapidly with respect to
$n$, even for polymers which have metallic ground states. For the
purpose of illustration we consider two model polymers namely chains
consisting of: (a) carbon atoms (henceforth C-chain), and (b) alternating
boron and nitrogen atoms (henceforth BN-chain). A C-chain consisting
of uniformly spaced atoms will be metallic, which, as per Peierls
theorem\cite{peierls}, is not allowed. Therefore, such a system is
expected to dimerize leading to an insulating ground state\cite{peierls}.
On the other hand Peierls theorem is not applicable to the BN-chain,
which is a band insulator and isolelectronic with the C-chain for
a two-atom unit cell. In an earlier from our group, we had studied
the ground state geometry of C- and BN-chains using a fully \emph{ab
initio} methodology both at the RHF and the correlated levels, and
concluded that C-chain does indeed exhibit dimerization, while the
BN chain prefers the uniform geometry\cite{ayjamal}. We explore the
ground state geometries of these two systems using our code. In order
to take care of the dangling bonds, we terminate the ends of oligomers
of uniform C-chain and the BN chain with two hydrogen atoms on the
each end. The dimerized C-chain consisting of alternating single and
triple bonds, on the other hand, is terminated by one hydrogen atom
on the each end.

\begin{figure}
\includegraphics[scale=0.5]{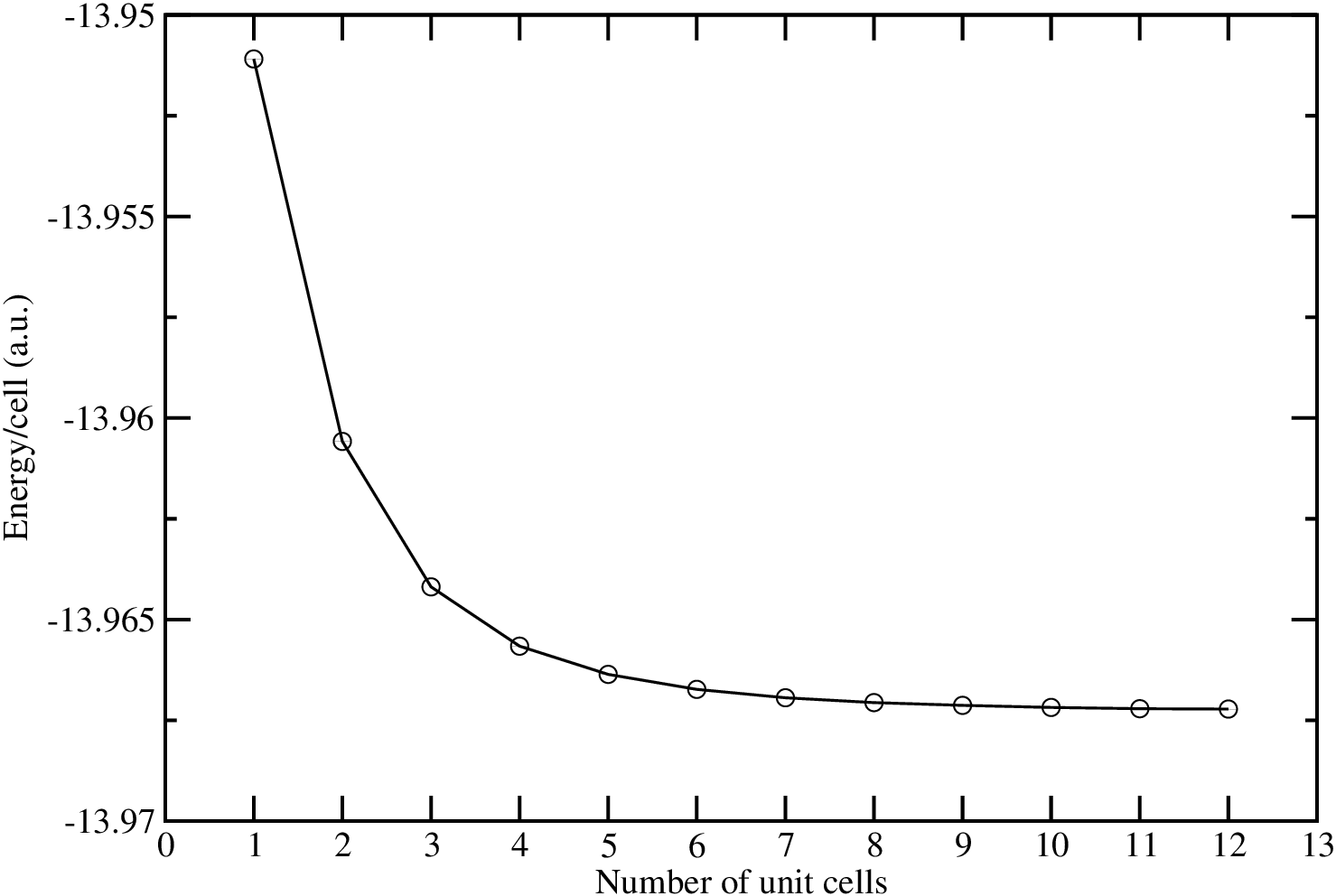}

\caption{Energy per two-atom unit cell, of an undimerized C-chain, plotted
as a function of the number of unit cells. The C-C bond length was
taken to be $1.297\textrm{\AA}$.}

\label{fig:en-c-chain}
\end{figure}

\begin{figure}[H]
\includegraphics[scale=0.5]{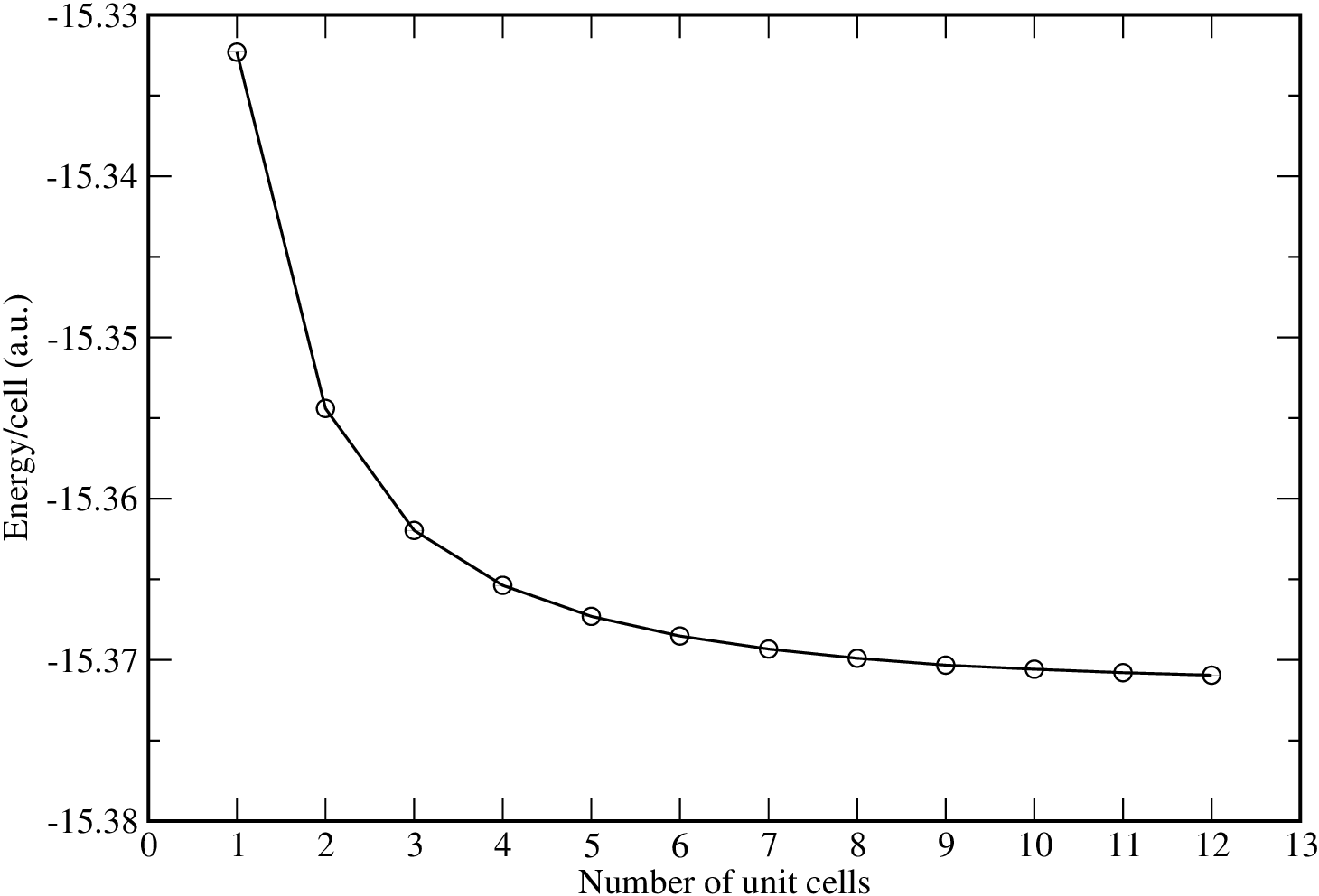}

\caption{Energy per unit cell of boron-nitrogen chain, plotted as a function
of the number of unit cells. The B-N bond length was taken to be $1.360$$\textrm{\AA}$.}
\label{fig:en-bn-chain}
\end{figure}

First we examine the convergence of $E_{cell}$ obtained using Eq.
\ref{eq:ecell} with respect to the number of unit cells $n$. In
Figs. \ref{fig:en-c-chain} and \ref{fig:en-bn-chain} we plot $\Delta E(n)$
as a function of $n$, for uniform C- and BN-chains, respectively.
In both the cases convergence with respect to $n$, for the two-atom
unit cells, is quite rapid, and for $n=10$ the bulk limit has been
achieved to reasonable accuracy. This is quite remarkable because
the C-chain considered for this calculation is metallic because of
uniformly placed atoms. The convergence is even more rapid for C-chains
with dimerized geometry.

Results of our calculations are summarized in tables \ref{tab:uniform-chain}and
\ref{tab:dimerized-chain}. From these tables, the following trends
are obvious: (a) CNDO/2 and INDO optimized geometries in all the cases
are in good agreement with each other, and (b) optimized bond lengths
obtained here are slightly larger than those obtained using the \emph{ab
initio} RHF method\cite{ayjamal}. 

Additionally, the condensation energy per atom of the C-chain, defined
as the difference in $E_{cell}$ per atom of the optimized geometries
in the uniform and dimerized configurations, are obtained to be 11.1
mHartrees/atom (CNDO/2), and 11.3 mHartrees/atom (INDO). These numbers
are in reasonable agreement with the corresponding \emph{ab initio}
RHF value of 7.8 mHartrees/atom\cite{ayjamal}.

\begin{table}
\caption{Comparison of our CNDO/2 and INDO geometries for uniform C-chain,
and the BN chain, with our earlier \emph{ab initio} RHF results\cite{ayjamal}.}

\begin{tabular}{|c|c|c|}
\hline 
Calculation & \multicolumn{2}{c|}{Bond Length (\AA)}\tabularnewline
\hline
\hline 
 & C-Chain & BN-Chain\tabularnewline
\hline 
This Work (CNDO/2) & 1.297 & 1.360\tabularnewline
\hline 
This Work (INDO) & 1.300 & 1.362\tabularnewline
\hline 
Abdurahman \emph{et al.\cite{ayjamal}} & 1.251 & 1.287\tabularnewline
\hline
\end{tabular}\label{tab:uniform-chain}
\end{table}

\begin{table}
\caption{Comparison of our CNDO/2 and INDO geometries obtained for the dimerized
C-chain, with our earlier \emph{ab initio} RHF results\cite{ayjamal}}

\begin{tabular}{|c|c|c|}
\hline 
Calculation & $r_{single}$(\AA) & $r_{triple}$(\AA)\tabularnewline
\hline
\hline 
This work (CNDO/2) & 1.390 & 1.231\tabularnewline
\hline 
This work (INDO) & 1.390 & 1.227\tabularnewline
\hline 
Abdurahman \emph{et al.\cite{ayjamal}} & 1.360 & 1.174\tabularnewline
\hline
\end{tabular}\label{tab:dimerized-chain}

\end{table}

\subsection{Optical Absorption in Fullerene C$_{60}$}

Since the discovery of the C$_{60}$ in 1985\cite{c60-exp}, the field
of the electronic structure and optical properties of fullerenes has
become one of the foremost research topics these days\cite{orlandi-review}.
Therefore, as the last application of our code in this paper, we present
the results of linear optical absorption calculations in C$_{60}$,
at the RHF level. For these calculations we utilized the same geometry
of Shibuya and Yoshitani\cite{shibu-yoshi}, as was used for total
energy calculations presented in section \ref{sub:molecules}. In
the CNDO/INDO models, with four basis functions per carbon atom, C$_{60}$
has 120 occupied and 120 unoccupied orbitals, with the ground state
being a closed shell with the $A_{g}$ symmetry. The HOMO/LUMO exhibit
nearly $\pi/\pi^{*}$ character, with a five-fold degenerate HOMO
($h_{u}$) and a three-fold degenerate LUMO ($t_{1u}$). Our HOMO-LUMO
gap of 9.23 eV for the INDO calcualtions is in perfect agreement with
that reported by Shibuya and Yoshitani\cite{shibu-yoshi}. 

\begin{figure}[H]
\includegraphics[width=12cm]{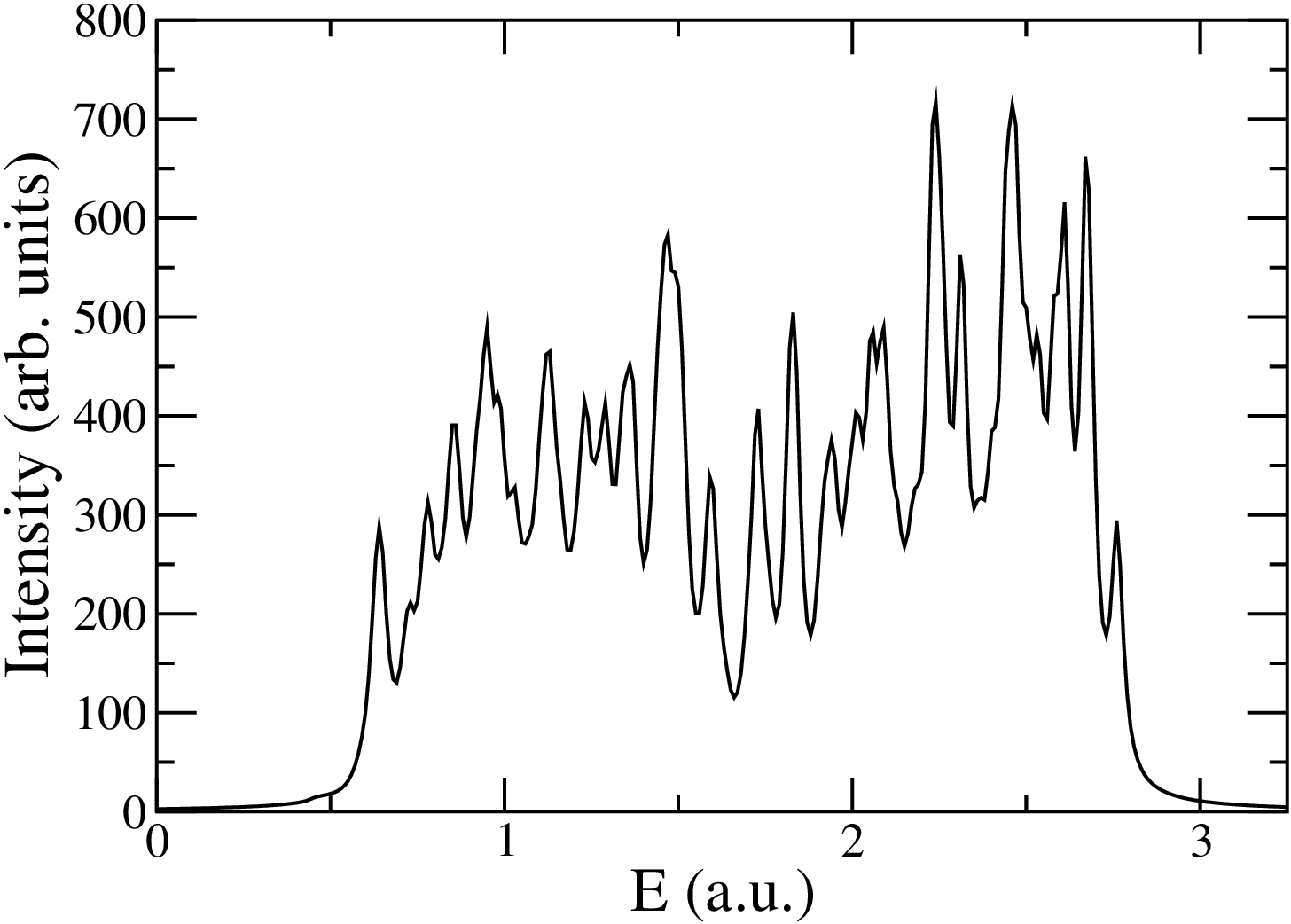}

\caption{Linear absorption spectrum of C$_{60}$, obtained from RHF calculations
using the INDO model, plotted as a function of the photon energy (in
atomic units). A line width of 0.02 a.u. was assumed. }

\label{fig:c60-spec}
\end{figure}

Next we present the linear optical absorption spectrum of C$_{60}$
computed by the INDO method under the electric-dipole approximation,
in Fig. \ref{fig:c60-spec}. Because the HOMO and LUMO orbitals have
the same inversion symmetry (\emph{ungerade}), the HOMO$\rightarrow$LUMO
transition is dipole forbidden leading to negligible absorption intensity
in the low-energy regions, in complete agreement with the experiments\cite{orlandi-review}.
We have intentionally plotted the spectrum with a relatively small
line width to emphasize the fact that a number of transitions among
various orbitals contribute to the linear absorption. Qualitative
features of our computed spectrum, namely the occurrence of two broad
bands with a number of subpeaks in the spectrum, are in good agreement
with other theoretical calculations\cite{c60-optics-theory}. As far
as the quantitative comparison with the experiments is concerned,
it is a well-known fact that the HF method overestimates the energy
gaps significantly. Therefore, in future works we intend to carry
out various levels of CI calculations to investigate the influence
of electron-correlation effects on the linear absorption in C$_{60}$.

\section{Conclusions and Future Directions}

\label{sec:conclusions}

In this paper we have described our Fortran 90 program which solves
the HF equations for both the closed- and open-shell molecular systems
using the semiempirical CNDO/2 and INDO models. To demonstrate the
correctness of our approach, we presented numerous test calculations
on molecular systems for which CNDO/INDO results are known, and obtained
essentially exact agreement. Additionally, we presented results on
systems such as clusters, fullerene, and polymers to demonstrate the
wide utility of our present program. The reason behind developing
the present program is twofold: (a) to develop a code in a modern
language such as Fortran 90 which can carry out dynamic array allocation
and thus free the user from specifying and changing array sizes, and
(b) to provide an open software which will be widely available to
users which they can use and modify as per their needs. One could
write programs to perform a change of basis on the Hamiltonian matrix
elements from the basis set AO representation to the MO representation,
and use the transformed Hamiltonian to perform correlated CI calculations.
Additionally, one could also introduce an electric-field in the Hamiltonian
to perform finite-field calculations to compute quantities such as
static polarizabilities of various orders. 

The present version of our code is restricted to first-row atoms using
the INDO method and up to the second-row elements using the CNDO/2
approach. It will be extremely desirable to extend these methods to
elements further in the periodic table, preferably up to the transition
metals. However, there are several versions of these models available
for heavier elements such as the s-p-d INDO, ZINDO, and other methods\cite{zindo}.
Therefore, one could implement these methods in the present code which
will allow the user to perform both INDO and CNDO/2 calculations on
elments of second-row and beyond. 

Work along those directions is continuing in our group, and results
will be published as and when they become available. 

\begin{ack}
Authors gratefully acknowledge a visit to Professor P. Fulde's group
in Max-Planck-Institut f\"ur Physik Komplexer Systeme, Dresden where
a part of this work was done. Additionally, we are also thankful to
Professors K. Dharamvir and V. K. Jindal for providing us with their
computer program for generating the atomic coordinates of C$_{60}$.
\end{ack}

\end{document}